\documentclass{article}
\usepackage{spconf,amsmath}
\usepackage{amsthm,amsmath,amssymb}
\usepackage{mathrsfs}
\usepackage{cite}
\usepackage{comment}
\usepackage{color}
\usepackage{booktabs}
\usepackage{graphicx}
\usepackage{subfigure}
\usepackage{tabularx}
\usepackage{caption}

\let\OLDthebibliography\thebibliography
\renewcommand\thebibliography[1]{
  \OLDthebibliography{#1}
  \setlength{\parskip}{0pt}
  \setlength{\itemsep}{0pt plus 0.3ex}
}

\begin{document}\sloppy

\def\x{{\mathbf x}}
\def\L{{\cal L}}

\title{An unsupervised optical flow estimation for LiDAR image sequences}
%
\name{Xuezhou Guo\textsuperscript{$\star$}, Xuhu Lin\textsuperscript{$\star$}, Lili Zhao, Zezhi Zhu, Jianwen Chen
\thanks{$\star$ Xuezhou Guo and Xuhu Lin contribute equally. © 2021 IEEE. Personal use of this material is permitted. Permission from IEEE must be obtained for all other uses, in any current or future media, including reprinting/republishing this material for advertising or promotional purposes, creating new collective works, for resale or redistribution to servers or lists, or reuse of any copyrighted component of this work in other works.}
}
\address{School of Information and Communication Engineering\\
University of Electronic Science and Technology of China\\ Chengdu 611731, China}

\maketitle

\begin{abstract}
In recent years, the LiDAR images, as a 2D compact representation of 3D LiDAR point clouds, are widely applied in various tasks, e.g., 3D semantic segmentation, LiDAR point cloud compression (PCC). Among these works, the optical flow estimation for LiDAR image sequences has become a key issue, especially for the motion estimation of the inter prediction in PCC. However, the existing optical flow estimation models are likely to be unreliable for LiDAR images. In this work, we first propose a light-weight flow estimation model for LiDAR image sequences. The key novelty of our method lies in two aspects. One is that for the different characteristics (with the spatial-variation feature distribution) of the LiDAR images w.r.t. the normal color images, we introduce the attention mechanism into our model to improve the quality of the estimated flow. The other one is that to tackle the lack of large-scale LiDAR-image annotations, we present an unsupervised method, which directly minimizes the inconsistency between the reference image and the reconstructed image based on the estimated optical flow. Extensive experimental results have shown that our proposed model outperforms other mainstream models on the KITTI dataset, with much fewer parameters.
\end{abstract}

\section{Introduction}
\label{sec:intro}
With the rapid development of Light detection and ranging (LiDAR) sensors, the captured sparse point clouds, which contain a set of 3D points representing the surrounding environment, have been widely used in various applications, such as, autonomous driving, augmented reality (AR), and drones. Recently, 2D LiDAR image sequences, which are generated by a projection from 3D LiDAR point clouds, are often encountered in many fields, e.g., the 3D semantic segmentation~\cite{squeezeseg,wu2019squeezesegv2,squeezeSegV3,RangeNet} and LiDAR point cloud compression (PCC)~\cite{Tu2016,Tu_ACCESS,TuMotion,Sun2020}. Among these works, some efforts on PCC aim to reduce the temporal redundancy of LiDAR point clouds, where the motion estimation becomes an essential issue. It is known that the scene flow~\cite{sceneflow} is a widely-used format of 3D motion field for LiDAR point clouds (e.g.,~\cite{pc_sceneflow1,pc_sceneflow2,pointflownet,flownet3d}). However, the scene flow estimation is not applicable for 2D LiDAR image sequences. Thus, an effective optical flow estimation method for LiDAR image sequences is needed.

\begin{figure}[t]
	\centering
	\subfigure[]{
		\begin{minipage}[t]{5cm}
			\includegraphics[width=4.8cm]{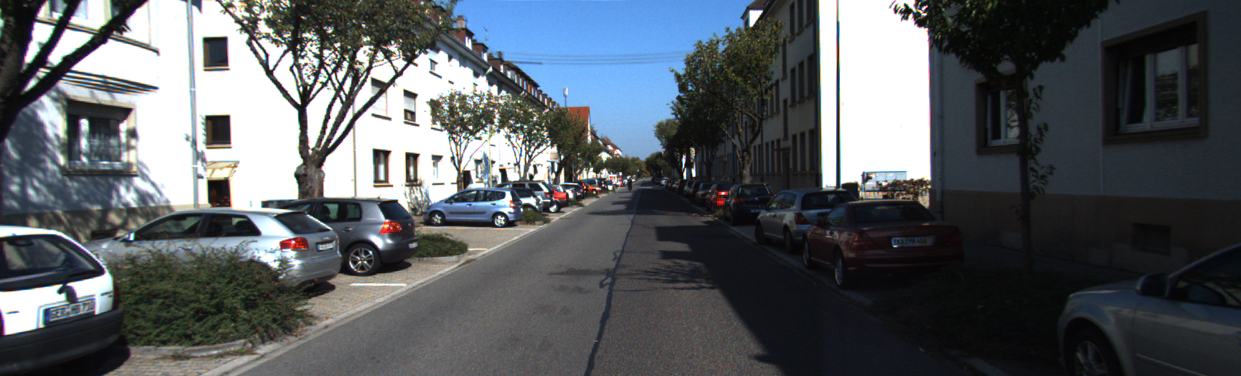}
		\end{minipage}
	}
	\subfigure[]{
		\begin{minipage}[t]{3cm}
			\includegraphics[width=3.1cm]{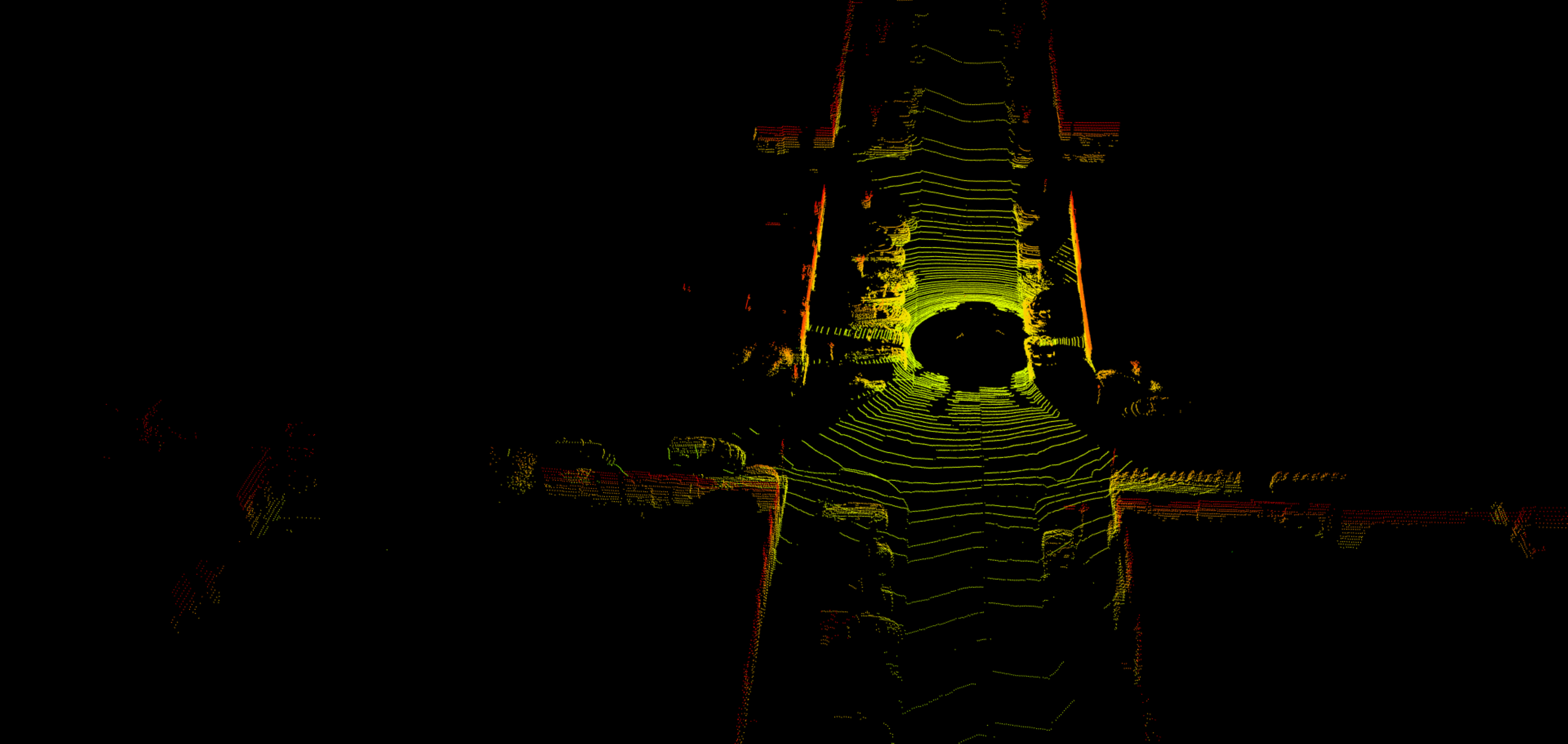}
		\end{minipage}
	}
	\subfigure[]{
		\begin{minipage}[t]{8.5cm}
			\includegraphics[width=8.4cm]{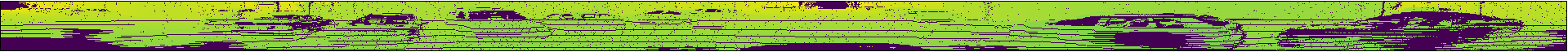}
		\end{minipage}
	}
	\caption{\textbf{Demonstration of one frame in the KITTI dataset.} (a) The original RGB image just for reference; (b) The LiDAR point cloud; (c) The corresponding LiDAR image.} 
	\label{fig:1}
\end{figure}

Recent years have witnessed remarkable success in applying CNNs for the optical flow estimation (e.g.,~\cite{dosovitskiy2015flownet,ilg2017flownet,sun2018pwc}). Among these methods, PWC-Net~\cite{sun2018pwc} is one of the state-of-the-art optical flow estimation algorithms. It exploits a pyramid network to extract features from the normal color images, and then constructs a cost volume as the measurement for the difference between warped and original features. It seems straightforward to naively feed LiDAR image sequences into the existing optical flow estimation models. However, directly applying these models leads to unsatisfactory results. This can be explained by the fact that the intrinsic spatial characteristics of LiDAR images are different from that of the normal color images, as shown in Fig.~\ref{fig:1}. Unlike the normal color images, the feature distribution of LiDAR images varies greatly at different pixel locations~\cite{squeezeSegV3}. Therefore, it is not effective to only use the standard convolutions for extracting the features of LiDAR images. 

\begin{figure*}
	\centering
	\includegraphics[width=18cm]{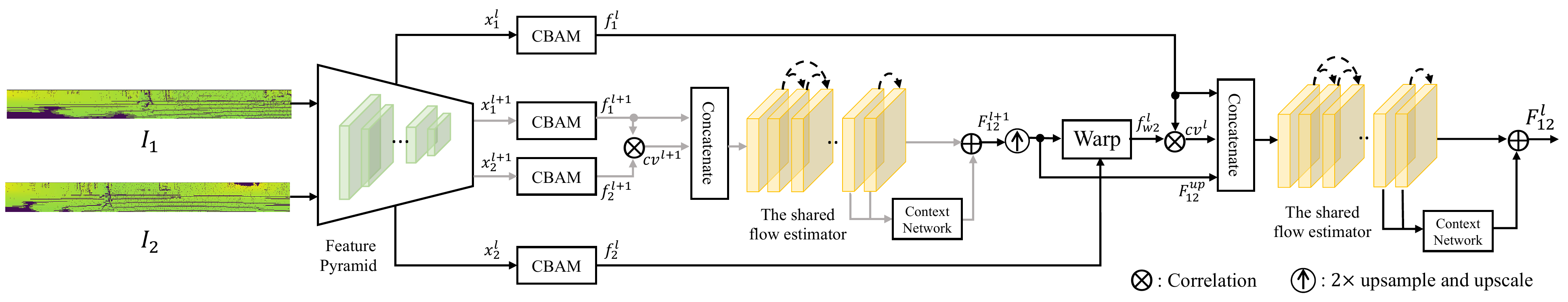}
		\caption{\textbf{The architecture of our proposed optical flow estimation model for LiDAR images.} It contains three main parts: the feature pyramid with CBAM, a shared flow estimator for all levels and the context network. For simplicity, we only show the pipeline of the top two levels in different line colors. The remaining modules have the same structure.}
	\label{network}
\end{figure*}

Fortunately, the attention mechanism has shown its potential to tackle this problem. Among the existing state-of-the-art algorithms about the attention mechanism, a Convolutional Block Attention Module (CBAM)~\cite{cbam} is proposed to adaptively refine features along the channel and spatial dimensions. The channel attention determines which features are more meaningful, and the spatial attention focuses on which portions of the image are more informative.

In this paper, we first propose a light-weight unsupervised flow estimation model for LiDAR image sequences. In general, the main pipeline of our work is based on PWC-Net~\cite{sun2018pwc}. Specifically, the LiDAR point clouds are firstly projected into a set of 2D LiDAR images as the inputs of the proposed model. Then, the `coarse' features of the input LiDAR images are extracted by a pyramid network, which is followed by a Convolutional Block Attention Module (CBAM)~\cite{cbam}. After that, the `refined' features will be fed into the subsequent flow estimator. Although the CBAM introduces additional parameters, by sharing the same flow estimator for all levels as \cite{liu2020learning}, our proposed model is with a smaller size than PWC-Net. Finally, a sub-network takes the last two outputs of the flow estimator as inputs and refines the estimated optical flow by enlarging the receptive field. 

The main contributions of this work are: \emph{(i)} To the best of our knowledge, we first propose an effective flow estimation model for LiDAR image sequences, which is based on the unsupervised learning; \emph{(ii)} To better handle the inherited spatial characteristics of the LiDAR images, we integrate the attention mechanism to improve the capacity of feature extraction for the standard convolutions;  \emph{(iii)} The proposed method achieves a comparable performance w.r.t. the mainstream optical flow estimation methods, but with fewer parameters.

\section{The Proposed Method}
Given two frames $I_{1}, I_{2}$ of the LiDAR image sequences, our goal is to estimate the forward optical flow $F_{12}$. In this section, we will first illustrate the projection method (i.e., the transformation from 3D point clouds to 2D LiDAR image sequences) for the data pre-processing. Then, various stages of the proposed framework will be described in details. 
 
\subsection{The Projection from the 3D LiDAR Point Clouds into the 2D LiDAR Image Sequence}
\label{projection}
As computed in~\cite{RangeNet}, the sparse LiDAR point clouds are firstly projected into a set of 2D LiDAR images (i.e., a LiDAR image sequence). For each point $(x, y, z)$ of LiDAR point clouds, a corresponding pixel $(u, v)$ of the LiDAR image sequence can be available as 
\begin{equation}
\left(\begin{array}{l}
u \\
v
\end{array}\right)=\left(\begin{array}{c}
0.5 \left[1-\arctan (y, x) \pi^{-1}\right] w \\
{\left[1-\left(\arcsin \left(z, \rho^{-1}\right)+f_{\text {down }}\right) f^{-1}\right] h}
\end{array}\right),
\end{equation}
where $w$ and $h$ represent the width and height of the projected LiDAR image, $\rho$ denotes the range from each point to the sensor's center as $\rho=\sqrt{x^{2}+y^{2}+z^{2}}$, and $f=\left|f_{\text {down }}\right|+\left|f_{u p}\right|$ is the vertical field-of-view (FOV) of the LiDAR sensor.

\subsection{The Optical Flow Estimation Model}
The main pipeline of our proposed model is illustrated in Fig.~\ref{network}. This model contains the following components: the feature pyramid network, the Convolutional Block Attention Module (CBAM), the cost volume, the optical flow estimator and the context network. We will discuss these sub-modules respectively.

\vspace{0.3em}
 \noindent \textbf{Feature Pyramid Network.} 
At the first stage, for two consecutive frames of LiDAR images $I_{1}$ and $I_{2}$, a seven-level pyramid network is adopted to generate their feature representations. From the 1st to the 7th level, the numbers of feature channels are $1$, $16$, $32$, $64$, $96$, $128$, $192$, respectively. Each level contains two convolutional layers followed by a Leaky ReLu function. At the $l^{th}$ level, the outputs of the pyramid network are the `coarse' feature representations $x^{l}_{1}, x^{l}_{2}$ for $I_{1}$ and $I_{2}$, respectively. Next, $x^{l}_{1}$ and $x^{l}_{2}$ will be refined by the CBAM.

\begin{figure}
	\centering
	\includegraphics[width=8cm]{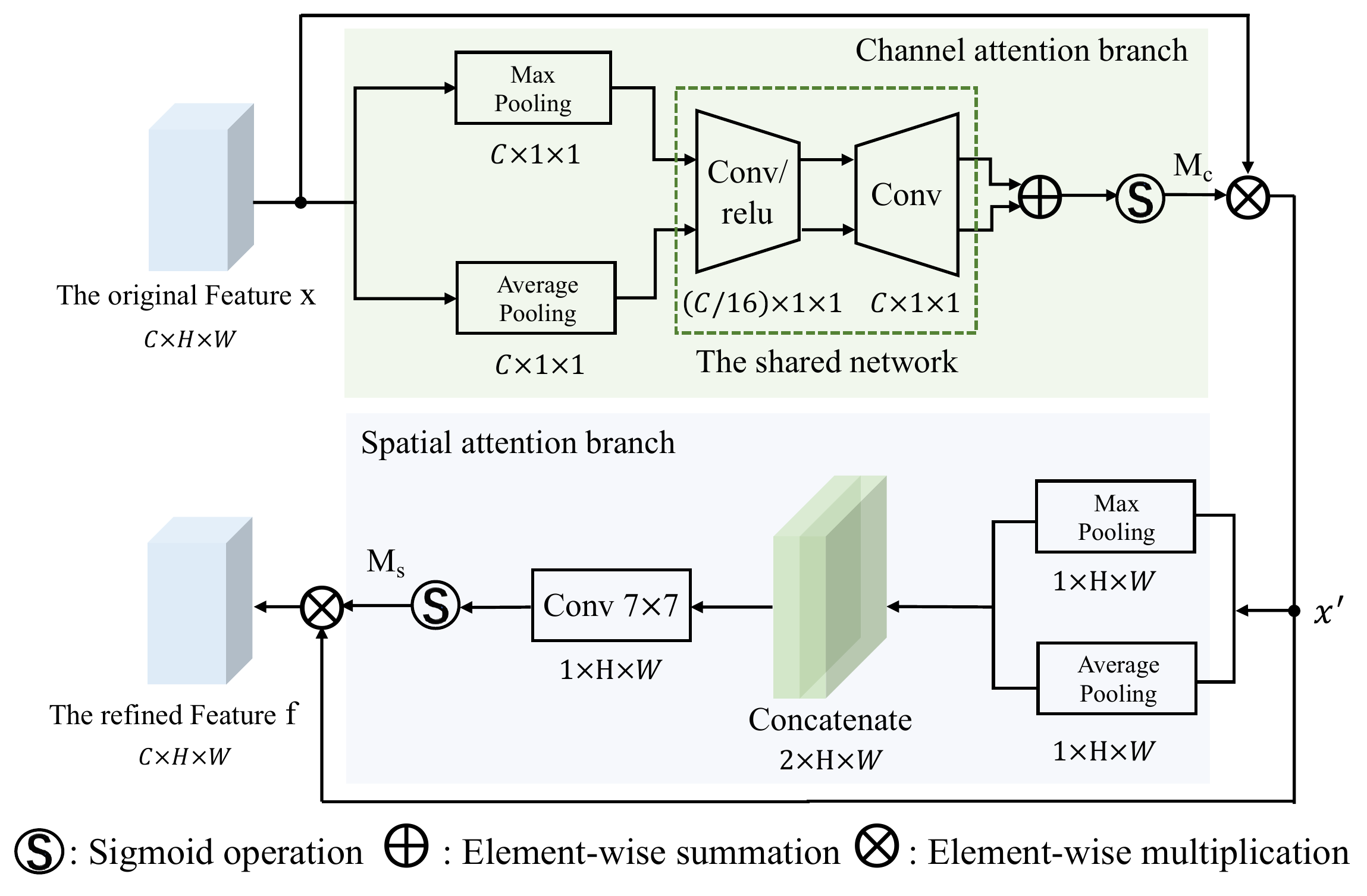}
	\caption{The structure of CBAM, which is proposed by~\cite{cbam}. It is embedded into our model to refine the extracted features.}
	\label{CBAM}
\end{figure}

\vspace{0.3em}
\noindent \textbf{Convolutional Block Attention Module.}
 The Convolutional Block Attention Module (CBAM) can adaptively select important features along the channel and spatial dimension. The structure of the CBAM is shown in Fig.~\ref{CBAM}. There are two main branches. Suppose that its input feature map is $x\in \mathbb{R}^{C \times H \times W}$, where $C$, $H$, $W$ denote the number of channels, the height, and the width of 3D features, respectively. 
 
 Firstly, at the channel attention branch as shown in Fig.~\ref{CBAM} (top), the CBAM learns a channel attention map $M_{c}\in \mathbb{R}^{C \times 1 \times 1}$ to refine $x$. The intermediate feature map generated in this branch is represented as $x^{\prime}$. Then, at the spatial attention branch as shown in Fig.~\ref{CBAM} (bottom), $x^{\prime}$ is fed as the input, and the spatial attention map $M_{s}\in \mathbb{R}^{1 \times H \times W}$ is learned to refine features $f$ along the spatial dimension, which is the final desired feature map. This process can be formulated as 
 \begin{equation}
 \begin{aligned}
 x^{\prime} =  x \otimes M_{c}(x), \\
 f =  x^{\prime} \otimes M_{s}(x^{\prime}),
 \end{aligned}
 \end{equation}
 where $\otimes$ denotes the element-wise multiplication. For the `coarse' feature representations $x^{l}_{1}, x^{l}_{2}$ from the pyramid network at the $l^{th}$ level, their refined features, denoted as $f^{l}_{1},f^{l}_{2}$, are used for the subsequent flow estimation process.

\vspace{0.3em}
\noindent \textbf{Cost Volume.}
The cost volume is widely used in the optical flow estimation tasks~\cite{ilg2017flownet,8100098}. It measures how closely a pixel's feature matches up with its corresponding pixel's feature in the next frame. Specifically, at the $l^{th}$ level, the obtained flow from the $l+1^{th}$ level, $F^{l+1}_{12}$, is up-sampled and up-scaled to the current level, $F^{up}_{12}$. It warps with $f^{l}_{2}$ towards the first image, and the warped feature map $f^{l}_{w2}$ will be obtained. After that, the cost volume $cv^{l}$ can be calculated as the correlation between $f^{l}_{1}$ and $f^{l}_{w2}$. For the top level, there is no optical flow from the upper level, so the cost volume is just the correlation between $f^{l}_{1}$ and $f^{l}_{2}$. 

 \vspace{0.3em}
\noindent \textbf{Optical Flow Estimator.}
The concatenation of $F^{up}_{12}$, $f^{l}_{1}$ and $cv^{l}$ is used as the input of the flow estimator. The flow estimator is a six-layer convolutional neural network (CNN). The numbers of the output channels for each layer are $128$, $128$, $96$, $64$, $32$, $2$, respectively. To reduce the computational complexity incurred by the introduced attention mechanism, we follow the architecture of the flow estimator adopted in~\cite{liu2020learning}. In specific, different from PWC-Net~\cite{sun2018pwc}, the inputs of each convolutional layer are only the outputs of the previous two layers. Besides, the flow estimators at different levels of the pyramid network share the same parameters, while PWC-Net is not in this case.

 \vspace{0.3em}
\noindent \textbf{Context Network.}
After obtaining the predicted flow, a context network is used for refinement. The context network is a CNN implemented by the dilated convolutions, which can enlarge the receptive field to further extract more contextual information~\cite{sun2018pwc}. The dilation constants for each level are set to $1$, $2$, $4$, $8$, $16$, $1$, $1$, respectively. The inputs of this sub-network are the outputs of the last two layers of the flow estimator. Finally, the refined flow $F_{12}^l$ is obtained by the context network.

\subsection{The Unsupervised Learning Strategy}
\label{loss}
There is no ground-truth optical flow for LiDAR image sequences. Therefore, our model is trained in an unsupervised way. Motivated by the work~\cite{unsuper}, we adopt the loss function designed for unsupervised optical flow learning tasks. It is to measure the difference between the reference image and the reconstructed one. The reconstructed image is obtained by back-warping the target image towards the reference image based on the estimated flow. Specifically, for the two input LiDAR images $I_{1}$ and $I_{2}$, at the $l^{th}$ level, $I_{1}$ and $I_{2}$ are down-sampled to match with the estimated flow $F_{12}^{l}$. The sampled images are denoted as $I_{1}^{l}$ and $I_{2}^{l}$. Then the reconstructed image $I_{r}^{l}$ can be calculated as 
\begin{equation}
I_{r}^{l}= backwarp(I_{2}^{l}, F_{12}^{l}).
\label{recons}
\end{equation} 

\subsection{Implementation Details}
\label{details}

\noindent 
\textbf{The Dataset.} 
We conduct the experiments on the KITTI dataset~\cite{geiger2012we}. It provides the odometry benchmark, which contains 22 sequences of the point clouds collected by the LiDAR sensors, covering diverse traffic scenes. For each sequence, we project the point clouds into LiDAR image sequences using the method illustrated in Section~\ref{projection}, as~\cite{squeezeseg,wu2019squeezesegv2,squeezeSegV3,RangeNet,Sun2020}. Then, we divide each three consecutive LiDAR images into a triplet. Specifically, 16 sequences are selected as the training set, while three sequences are chosen as the validation set. The remaining sequences are used as the test set. For each triplet, two range images are used as the inputs to predict the optical flow. 

\vspace{0.3em}
\noindent 
\textbf{The Loss Function.}
Based on the illustration of the unsupervised loss function described in Section~\ref{loss}, different terms are used for the training and fine-tuning of our model. For the training process, following the main pipeline of the multi-scale loss proposed in~\cite{dosovitskiy2015flownet}, we use the weighted sum of losses from seven pyramid levels to guide the optical flow learning. The training loss can be formulated as:
\begin{equation}
\mathcal{L}_{t} = \sum_{l=1}^{7} \alpha_{l} \|I_{r}^{l} - I_{1}^{l} \|_{2},
\end{equation}
where $\alpha_{1}$ to $\alpha_{7}$ are set as $0.3$, $0.06$, $0.08$, $0.1$, $0.12$, $0.14$, $0.2$, respectively, and $I_{r}^{l}$ is from Equation (\ref{recons}).
For the fine-tuning process, recent studies~\cite{liu2019selflow, meister2018unflow} use an occlusion map in the loss function, so as to exclude the effects of occluded pixels. Inspired by this, in the reconstruction process, we use a binary map $M^{l}_{b}$ generated from $I^{l}_{1}$ to indicate the existence of a point in the space. Modified from Equation (\ref{recons}), the reconstructed image is represented as
\begin{equation}
I^{l}_{b}= backwarp(I_{2}^{l}, F_{12}^{l}) \times M^{l}_{b}.
\end{equation} 

Then, the fine-tuning loss function is formulated as
\begin{equation}
\label{ftloss}
\mathcal{L}_{ft} = \sum_{l=1}^{7} \alpha_{l} \|I^{l}_{b} - I_{1}^{l} \|_{2} + \gamma\|\theta\|_{2},
\end{equation} 
where $\theta$ is the learnable weight of this model. In our experiments, we set the regularization parameter $\gamma$ to $1e-6$.

\vspace{0.3em}
\noindent 
\textbf{The Training Strategy.}
Firstly, the proposed model is trained from scratch for 60 epochs, with a batch size of 4 and the initial learning rate is 1e-4. The learning rate will decay by 0.1 for every 20 epochs. We adopt the ADAM optimizer with $\beta_{1}=0.9$, $\beta_{2}=0.999$ and $\epsilon=1e-7$. Next, the network is fine-tuned with a batch size of 1, using the modified loss function, Equation (\ref{ftloss}), for another 40 epochs. The initial learning rate is set to 0.5e-4, and it will decay by 0.5 when the validation loss has stopped decreasing for 4 epochs. The ADAM optimizer has the same parameter setting for both training and fine-tuning processes.

\section{Experimental Results}
We evaluate the performance of our model on the test set. Several mainstream flow estimation models designed for the normal color images, are compared quantitatively. The experiments are conducted using an Intel Core i7-7700K CPU with a GTX 1080Ti GPU.

\subsection{Evaluation Metrics}
It is known that the endpoint error (EPE)~\cite{unsuper} is used as the metric for the supervised optical flow learning tasks. However, there is no ground-truth flow annotations for the LiDAR image sequences. Therefore, an `indirect' metric is used for evaluation in this paper, i.e., Manhattan distance (i.e., L1 norm), measuring the error between the reference image and the reconstructed one. The reconstructed image is obtained as Equation (\ref{recons}).

\subsection{Performance Evaluation}
\begin{figure}[t]
 \centering
 \includegraphics[width=8cm]{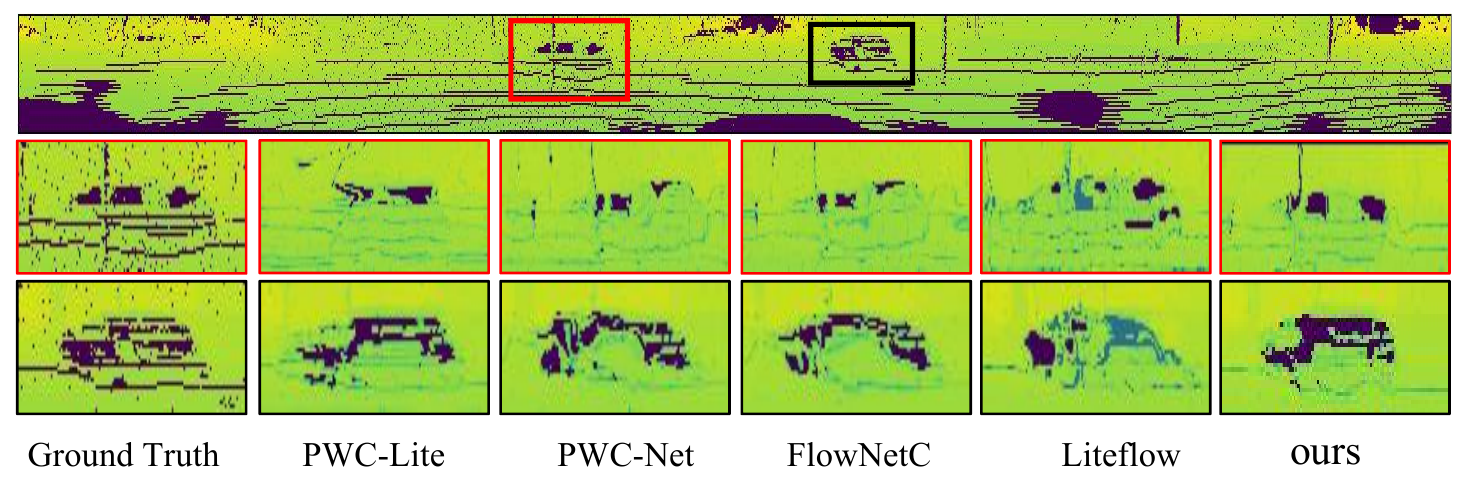}
 \caption{Visual comparison for the reconstructed images. Our model delivers better perceptual quality than other methods.}
 \label{resultgraph}
\end{figure}

 From Table \ref{result}, it can be observed that our proposed method outperforms other methods in terms of the inference loss, especially with an improvement of 50$\%$ than the original PWC-Net. This result shows the effectiveness of introducing the attention mechanism into the feature extraction. Besides, the results indicate our method has fewer parameters. Note that compared with PWC-Lite, our method slightly increase the model size, but we deliver a better flow estimation. Moreover, we also provide the results of our model with fine-tuning or not, for showing the effectiveness of adding the binary map in the loss function. Fig.~\ref{resultgraph} shows the comparison between the ground-truth and reconstructed frame using different flow estimation models. We select two frames from the test set, and set the first image as the ground truth. The reconstructed frame is obtained by back-warping the second frame based on the estimated optical flow. It can be found that compared with other methods, our proposed model can reconstruct a more accurate LiDAR image, especially for the object's size and shape. These results all demonstrate the effectiveness of our flow estimation method.

\begin{table}[t]
    \footnotesize
	\caption{\textbf{Quantitative comparisons for the optical flow methods on the KITTI test dataset.} `c' means the model without fine-tuning and `ft' means the model with fine-tuning. }
	\centering
	\begin{tabular}{ccccc}
		\toprule
		&Method & \# Param. & L1 loss&\\
		\midrule
		&PWC-Net~\cite{sun2018pwc} & 9.37M & 3.1848&\\
		&PWC-Lite~\cite{liu2019selflow} & \textbf{2.24M} & 2.7664&\\
		&FlowNetC~\cite{ilg2017flownet} & 39.18M & 2.5031&\\
		 &LiteFlowNet~\cite{hui18liteflownet} & 5.38M & 4.3231&\\
		 \midrule
		 &Ours(c) & 2.25M & 1.9870&\\
		 &Ours(ft) & 2.25M & \textbf{1.7568}&\\
		\bottomrule
	\end{tabular}
\label{result}
\end{table}
\vspace{-1em}

\section{Conclusion}
In this paper, a novel flow estimation method is firstly proposed for the LiDAR image sequences, which exploits the attention mechanism strategy to further improve the quality of flow estimation. Moreover, to address the issue about the lack of the annotations of LiDAR images, the unsupervised learning strategy is introduced by minimizing the error between the reference image and the warped image obtained by the estimated optical flow. Extensive simulation results on the KITTI dataset have clearly shown that for the optical flow estimation of the LiDAR image sequences, our proposed model outperforms the existing flow estimation models, even with much fewer parameters.

 \section{ACKNOWLEDGMENT}
 This work is supported by the Sichuan Science and Technology Program under grant No. 2018JY0610.
%


\bibliographystyle{IEEEbib}
\bibliography{icip2021template1}

\end{document}